\documentclass[aps, prd, twocolumn, lengthcheck, superscriptaddress, 
nofootinbib]{revtex4-1}

\usepackage{epsfig}
\usepackage[usenames]{color}
\usepackage{graphicx}
\usepackage{amsmath}
\usepackage{epstopdf}

\def\bi{\bibitem}

\def\la{\langle}\def\ra{\rangle}
\def\be{\begin{eqnarray}}\def\ee{\end{eqnarray}}
\def\lsim{\mathrel{\rlap{\lower3pt\hbox{\hskip1pt$\sim$}}
     \raise1pt\hbox{$<$}}} 
\def\gsim{\mathrel{\rlap{\lower3pt\hbox{\hskip1pt$\sim$}}
     \raise1pt\hbox{$>$}}} 

\def\qgA{g^{O}_A}

\def\BGT{\cal{B}_{\rm GT}}

\allowdisplaybreaks

\begin{document}

\title{Anomaly-Induced Quenching of  ${g_A}$  in Nuclear Matter \\
and Impact on  Search for Neutrinoless $\beta\beta$ Decay}


\author{Mannque Rho}
\email{mannque.rho@ipht.fr}
\affiliation{Universit\'e Paris-Saclay, CNRS, CEA, Institut de Physique Th\'eorique, 91191, Gif-sur-Yvette, France }

\date{\today}

\begin{abstract}

How to disentangle the possible {\it genuine}  quenching of $g_A$ caused by scale anomaly of QCD parameterized by the scale-symmetry-breaking quenching factor $q_{ssb}$ from nuclear correlation effects is described. This is done by matching the Fermi-liquid fixed (FLFP) point theory  to the ``Extreme Single Particle (shell) Model" (acronym ESPM) in superallowed Gamow-Teller transitions in heavy doubly-magic shell nuclei. The  recently experimentally observed indication for $(1-q_{ssb})\neq 0$ -- that one might identify as ``fundamental quenching ({\it FQ})" -- in certain experiments seems to be alarmingly significant. I present arguments how symmetries hidden in the matter-free vacuum can emerge and suppress such {\it FQ}  in strong nuclear correlations. How to confirm or refute this observation is discussed in terms of the superallowed Gamow-Teller transition in the doubly-magic nucleus $^{100}$Sn and in the spectral shape in the multifold forbidden $\beta$ decay of $^{115}$In. 

\end{abstract}

\maketitle

\section{Introduction}
In \cite{gA-MR}, it was agued that what has been referred to as ``quenched  $g_A$" (denoted as $\qgA\simeq 1$ in what follows) observed in Gamow-Teller transitions in light nuclei~\cite{gA-review}  has little to do with {\it genuine} quenching of the axial-vector coupling constant measured  in free space, $g_A=1.276$.  It was suggested that what it represents is not a genuine  renormaiization of the axial coupling constant  in the effective field theory (EFT) space defined by the chiral scale $\lsim 4\pi f_\pi$  but the effect due to {\it full}  nuclear correlations driven by emerging scale symmetry.\footnote{What's commonly denoted as ``quenched $g_A$" in the literature is a misnomer as will be explained below.}   That it can be, mostly if not all,  due to an effective nuclear correlation mechanism was  arrived at in various models by many authors in the past as listed, e.g., in the reviews~\cite{gA-review}. To quote one early example out of many, in 1977 Les Houches Lectures, Wilkinson concluded, based on his analyses, that with the shell-model wave-functions for light nuclei $A\leq 21$ that include ``full mixing,"   the ``effective $g_{Ae}$" accounting for $\qgA=1$ comes out to be~\cite{wilkinson}
\be
g_{Ae}/g_A=1+(2\pm 6)\label{wilkinson}
\ee
implying that  there was nothing ``fundamental" in $\qgA$ going down to  $\sim$ 1 from the $g_A=1.276$ listed in the Particle Physics Booklet.  Indeed the modern high-power many-body calculations, such as  Quantum Monte Carlo calculations, in light nuclei requires no renormalized $g_A$, modulo possibly a few \% corrections coming from two-body (and more-body exchange) currents~\cite{wiringa}. 
 
The point stressed in \cite{gA-MR} was that  $g_{Ae}/g_A$ should be near 1 not just in light nuclei but also in heavy nuclei as well as highly dense nuclear matter {\it unless unsuspected quantum anomaly effects intervene}.  The situation in heavier nuclei as listed in the reviews~\cite{gA-review} looks  rather different, leaning toward  (\ref{wilkinson}) on the larger error bar side~\cite{gA-review},  hinting at possible density dependence.  This issue gets sharpened in the most recent experimental result in the superallowed Gamow-Teller transition in the doubly magic nucleus $^{100}$Sn~\cite{RIKEN}  where $g_{Ae}/g_A$ seems to be much less than 1. If the present data (and its interpretation) is correct, then there can be a big {\t FQ}. This is the issue I address here.
 
 In this short note, I discuss  there can in fact be a significant  renormalization of $g_A$ in nuclear medium {\it induced by the vacuum change by density} which I will call {\it genuine}  quenching of $g_A$. I will use the word {\it genuine} written in italic to be distinguished from what has been  referred in the literature to as  ``quenched $g_A$." My argument will rely on the notion of scale invariance in QCD put forward by Crewther~\cite{GD} invoking ``{\it genuine dilaton}."  If this argument is confirmed, it would drastically impact not only certain intrinsic properties of nuclear dynamics where pion-nuclear coupling constants are involved but also all nuclear responses to the weak axial current and perhaps more importantly the searches for going beyond the Standard Model (BSM) such as in neutrinoless double $\beta$ decay.
 
{ I will first describe what the solution to the early 1970's totally misunderstood issue ``$g_A$ puzzle" ($g_{Ae}/g_A=1$ in Wilkinson's  formula) could be,  what it implies in modern language how hidden (or spontaneously broken) scale symmetry manifests in nuclear medium and what ``fundamental" information vis-\`a-vis with the quenched $g_A$  could be involved in the future development for going beyond the Standard Model.}

What's involved is an up-to-date totally unexplored  novel mechanism  based on a symmetry hidden in QCD that seems to ``emerge" in strong nuclear correlations. {What is remarkable is that it seems to  permeate from nuclear-matter density $n_0\simeq 0.16$ fm$^{-3}$ to high densities $n_{star}\sim 7 n_0$ relevant to compact-star physics, and even beyond to what is referred to as ``dilaton-limit fixed point (DLFP)" $n_{\rm dlfp} \gsim 25 n_0$. The deviation of $\qgA$ from 1 observed in the $^{100}$Sn decay would, therefore, represent an anomaly in scale symmetry manifested, {\it not in the matter-free vacuum, but  in nuclear medium.}

\section{The ``{genuine dilaton (GD)}" and nuclear axial current}
I first describe how the scale (trace) anomaly of QCD can enter in the nuclear weak current. For this I adopt the notion of the ``genuine dialton (GD)" in QCD~\cite{CT,GD}.
The GD scheme is characterized by the existence of an infrared  fixed point (IRFP) $\alpha_{IR}$  at which both scale symmetry and chiral symmetry (in the chiral limit)  are realized in the Nambu-Goldstone (NG) mode,  populated by the massless NG bosons $\pi$ and dilaton $\sigma_d$ whose decay constants are non-zero. What is characteristic of this notion, crucially relevant,  is that it accommodates the massive nucleons $\psi$ and vector mesons $V_\mu$ at the IR fixed point.  I should mention here that there is  a new development -- referred to as conformal dilaton (CD) -- involving an IR fixed point structure accommodating massive hadrons that resembles the GD scenario,  possibly linking to the conformal window being discussed for large number of flavors relevant for dilatonic Higgs~\cite{CD, Zwicky}. Whether or not  and how these two schemes are related is not yet quite clear. In this paper, I will adopt the GD scheme although this scheme present in QCD (for $N_f\leq 3$)  is generally dismissed by those working on dilatonic Higgs theories for large $N_f$. It seems however feasible to justify the notion advocated here  in nuclear medium where scale symmetry ``emerges" from or is ``made visible" by strong nuclear correlations.

Given that what's involved is nuclear-matter density inaccessible from both lattice and perturbative QCD, our approach is inevitably anchored on an effective field theory. The relevant degrees of freedom that figure are, apart from the nucleons and NG bosons ($\pi$, $\sigma_d$),   the vector mesons $V=(\rho,\omega)$ as  flavor gauge fields. It therefore involves hidden local symmetry (HLS)~\cite{HLS,HLS-Komargodski}  and hidden scale symmetry (HSS)~\cite{GD,CT,CD,Zwicky}. These are the minimal number of degrees of freedom that  enable us to go  from normal nuclear matter to, involving a topology change at $n_{1/2}\sim (2-3)n_0 $, near the center of massive compact stars,  $n\sim (5-6) n_0$~\cite{MR-review}.

It turns out that an EFT with systematic power counting in both chiral and scale symmetries  (``CS"  for short) can be formulated~\cite{cata,Li-Ma-Rho} but in practice it has too many unknown parameters to fix, so it is not feasible at present to formulate an EFT as powerful as the standard chiral perturbation approach~\cite{bira}. Fortunately a highly powerful way to bypass the difficulty exists. It is to formulate the many-nucleon problem on the Fermi sphere.  {In renormalization-group (RG) approach to interacting fermions~\cite{shankar}, one can transform the mean fields of the leading CS-order effective Lagrangian with  hidden local symmetry and scale symmetry incorporated denoted as ${\cal L}_{\psi\chi {\rm HLS}}$ -- with $\psi$ standing for the nucleon fields,  $\chi$ for the ``conformal compensator field" for the dilaton $\chi=f_\chi e^{\sigma_d /f_\chi}$ and HLS for the vector fields -- to the Landau-Fermi-liquid (LFL) fixed point theory of many-nucleon systems. At the fixed point, one obtains the nuclear matter at its equilibrium~\cite{FR}.  Now with the nucleons put on a Fermi sphere, the LFL fixed point approximation corresponds to taking $1/\bar{N}$ to zero where $\bar{N}=k_F/(\Lambda_{fs} - k_F)$ with $\Lambda_{fs}$ the cut-off on top of the Fermi-surface measured with respect to the origin. It becomes more reliable as density increases.} As a bona-fide EFT, one can  do higher-order corrections in $1/\bar{N}$ in what's known as ``$V_{low K}$ RG-expansion" as done in finite nuclei. For the $g_A$  problem explicit $1/\bar{N}$ corrections are found to be unnecessary.

One can consider this approach as a DFT (density-functional theory) \`a la Hohenberg-Kohn theorem applied to nuclear matter. It can be thought of as an improved  version of Walecka's relativistic mean field theory of nuclear matter~\cite{walecka} with the refinement brought in by the hidden symmetries and the intrinsic density dependence~\cite{BR91} via the dilaton condensate $\la\chi\ra$.\footnote{The BR-scaling masses of the vector mesons account for higher meson-field terms consistently with the symmetries involved -- which are missed in nonlinear Walecka models.} The link to the Fermi-liquid structure of the Walecka's model has been discussed~\cite{matsui}.

\section{Quenching of  $g_A$ in G$n$EFT}
The LFL fixed point approach defined above, referred to as G$n$EFT in our work, is the framework that was exploited to address compact-star physics at high density in \cite{MR-review}. (Let me just mention that it has been highly successful for explaining  even the recent data from NICER/XMM-Newton~\cite{PCMstar}.)  Here I will focus entirely on the nuclear axial current.

The power of the G$n$EFT, even when drastically simplified,  has been that low-energy theorems, both iso-vector and iso-scalar, can be encoded in ${\cal L}_{\psi\chi {\rm HLS}}$-type Lagrangians and when applied to nuclear matter with suitable Ward identities, give very accurate results for nuclear electromagnetic (EM) response functions. { For instance, it has been shown~\cite{FR}  that for a quasiparticle on top of the Fermi surface, the  iso-scalar orbital gyromagnetic ratio $g^{0}_l=1$ -- with  the nucleon mass dropping \`a la  BR scaling at increasing density -- comes out to be consistent with the Kohn theorem~\cite{kohn-theorem} and the anomalous iso-vector gyromagnetic ratio $\delta g_l^{i}$, reproducing exactly the  Migdal formula~\cite{migdal} in terms of the Landau-Migdal parameter $F_1$,  agrees  exactly with the available Pb data.} To the best of the author's knowledge, there has not been as simple a derivation of, and with such a quantitative agreement with nature,  for these and other EM functions in (heavy) nuclei obtained in standard nuclear chiral perturbative (NchiPT) approach.  This same formalism will be applied to the axial-vector transitions in nuclei.

\subsection{$g_A^L$ as a Landau Fermi-liquid fixed-point quantity}

Consider the Landau quasiparticle propagating with near zero momentum on top of the Fermi sea in interaction with charged pionic and  vector-mesonic fluctuations on the surface. Via low-energy theorems,  such as  the Goldberger-Treiman relation, the coupling constant involved can be written with the axial coupling $g_A$. In the same LFL fixed-point approximation applied in the EM case, the ``effective" axial coupling of the quasiparticle on the Fermi surface can be calculated in the mean-field approximation as a Landau fixed-point quantity~\cite{FR}
\be
g_A^L=g_A q^{\rm Landau}_{snc}\simeq 1\label{FRgA}
\ee
with 
\be
q^{\rm Landau}_{snc}=(1- \frac 13 \Phi^\ast \bar{F}_1^\pi)^{-2}.
\ee
Here 
\be
\Phi^\ast=f_\pi^\ast/f_\pi
\ee
is the BR scaling~\cite{BR91} and $\bar{F}_1^\pi$ is the pionic contribution to the Landau mass. As defined, the factor $q^{\rm Landau}_{snc}$ is to capture the {\it complete} nuclear correlations (or full mixing in the sense of (\ref{wilkinson})). It can be unambiguously calculated and was found to be (with $g_A=1.276$)
\be
q^{\rm Landau}_{snc}=g_A^L/g_A \simeq 1/g_A\simeq 0.78.\label{Landau-snc}
\ee
This factor turns out to be  insensitive to density between $n=n_0/2$ and $n_0$. Note that the underlying assumption for (\ref{FRgA})  is that $N_c$ is large as in the Goldberger-Treiman relation in the vacuum and  $1/\bar{N}\to 0$. Thus the quenching of $g_A$ to 1 by the factor (\ref{Landau-snc}) is accounted for {\it entirely} by nuclear quasiparticle correlations on the Fermi surface. Now the question that was left un-clarified in \cite{FR} was: How is this $g_A^L\simeq1$  related to the quenching of $g_A$ in nuclear weak processes, in particular,  Gamow-Teller transitions in finite nuclei~\cite{gA-review}? To answer this question, look at the nuclear axial response function to the external weak field ${\cal W}_\mu$. What's involved is the iso-vector nuclear axial current $J_{5\mu}^a$ relevant in the EFT defined by the chiral scale.

At the classical level, the nuclear axial current coupling to the external weak field in the CS Lagrangian is scale-invariant. However there is an anomalous dimension contribution from the trace anomaly of QCD that enters nonperturbatively at the leading-order (LO) chiral-scale  perturbation expansion~\cite{CT}. The full LO axial current  is given by~\cite{GD} 
\be
J^{a\mu}_{ 5}=g_A q_{\rm ssb} \bar{\psi}\gamma^\mu\gamma_5 \frac{\tau^a}{2}\psi\label{AC}
\ee
{ where 
\be
q_{\rm ssb} (\beta^\prime, \chi)= c_A+(1-c_A)(\frac{\chi}{f_\chi})^{\beta^\prime}.\label{qssb}
\ee
If it were not for $q_{\rm ssb}$, the current (\ref{AC}) would be scale-invariant.  The scale-symmetry breaking enters in $q_{ssb}$ nonlinearly dependent on density.  While $c_A$ is a constant, possibly density-dependent in nuclear medium, the second term of $q_{ssb}$ carries the conformal compensator field $\chi$~\cite{GD}. In the matter-free vacuum,  $c_A$  could perhaps  be ``measurable" on lattice but in medium, can be accessed neither by lattice nor by perturbation. $\beta^\prime$ is the derivative of the $\beta (\alpha_s)$ at the IR fixed point
\be
\beta^\prime|_{\alpha_s=\alpha_{\rm IR}} 
\ee 
which is expected to be $> 0$ in the GD scheme but has not yet been measured on lattice for $N_f \leq 3$.  One sees that $\beta^\prime$ brings in an anomalous dimension, representing scale-symmetry explicit breaking, to the current operative in medium.
It is this quantity that could lead to a fundamental renormalization of the constant $g_A$.


\subsection{Accessing $q_{\rm ssb}$}

Now here is the most significant observation to make in the formula (\ref{qssb}).  

It is a nonlinear functional of $\chi$, apparently too complicated to treat in general. However in the problem concerned in nuclear matter, it can be simplified.  In the matter-free space, the vacuum expectation value (VeV) is $\la\chi\ra=f_\chi$, so if one ignores the fluctuating dilaton field that enters at higher loop orders justifiable  in the axial current,  one can set $q_{\rm ssb}=1$.  Then the  current remains scale-invariant.  There will then be no $\beta^\prime$ effect at the order involved.  However in nuclear matter,  $\la\chi\ra^\ast=f^\ast_\chi\neq f_\chi$ brings in scale symmetry breaking, both explicit  and spontaneous, hence bringing in the $\beta^\prime$ dependence.  
Applied to nuclear matter, one then has the density-dependent factor that I call ``anomaly-induced quenching"  ($AIQ$ for short)
\be
 q^\ast_{\rm ssb} = c_A +(1-c_A)(\Phi^\ast)^{\beta^\prime}\label{aiq}
\ee
where
\be
\Phi^\ast= f_\chi^\ast/f_\chi \simeq  f_\pi^\ast/f_\pi\label{Phi}
\ee
with  $\ast$ standing for density dependence.  The relation  $f_\chi^\ast/f_\chi \simeq f_\pi^\ast/f_\pi$ reflects the characteristic of the GD scheme, different from dilatonic Higgs models for large $N_c$~\cite{Appelquist}.

{\it The conclusion here is that within the G$n$EFT scheme,  $ q^\ast_{\rm ssb} < 1$ is the  fundamental quenching factor that can intervene in nuclei. It is a {\it genuine} quenching.  And it is the only {\it FQ} predicted in this G$n$EFT approach.}

 Note that this $AIQ$ was missing in the Goldberger-Treiman-type result in \cite{FR}.  There  $q^\ast_{\rm ssb} =1$ was set invoking the  LOSS (leading-order-scale-symmetric) approximation~\cite{MR-review}. 
 
 I should stress before going further that  the possible presence of the $AIQ$ -- and other $\beta^\prime$ effects -- has never been observed or even suspected in the literature  up to date. This article is the first to address the issue because of the important RIKEN data that if correct would drastically impact not only on the search for neutrinoless double-beta decay for going beyond the Standard Model but also on certain properties of nuclei and nuclear matter controlled by the combined scale and chiral symmetries.

\section{Mapping the Landau-Fermi-liquid fixed point approximation to the shell model}
In the absence of reliable full {\it ab initio} numerical treatments of nuclear many-body problems for complex nuclei -- apart from very light nuclei $A< 10$, how to translate the LFL fixed-point  result given above for nuclear matter to finite nuclei has not been  obvious. However there is one case, I argue, where this translation is feasible  and that is the supperallowed Gamow-Teller transitions in doubly-magic-shell nuclei that are heavy enough to be treated as a nuclear matter.

The LFL fixed-point limit in the Fermi-liquid system obtained above with $q_{\rm snc}^{\rm Landau}$ corresponds to  a Gamow-Teller transition undergoing on the Fermi surface with the kinematics $\omega\to 0$, $q/\omega\to 0$ -- where $(q,\omega)$ are the (momentum, energy) carried by the weak field. What is required for this limit to hold is that quasiparticle-quasihole bubble contributions entering at higher orders in $1/\bar{N}$ be suppressed. 

I now argue that this LFL fixed-point limit in the Fermi-liquid system can be mapped to the ``Extreme Single Particle (shell) Model (EPSM)"  in doubly-magic-shell nuclei. The most illustrative case is the transition in the $^{100}$Sn nucleus which has the proton and neutron shells completely filled at 50/50. There may be other nuclei of similar structure or perhaps even better but I will focus on this nucleus because it has the advantage  of being the heaviest nucleus with the equal magic shells that has also been studied extensively both experimentally and theoretically~\cite{GSI,FGG,RIKEN} with the results directly relevant to the issue concerned.

The process involved is a pure supperallowed GT transition of a proton ($\pi$) $\pi 0 g_{9/2}$ in the  completely filled orbital  into a neutron ($\nu$) in the empty spin-orbit partner, the $\nu 0 g_{7/2}$ orbital of  $^{100}$In. This offers the most favorable structure of the daughter state that is of a pure $(\nu g_{7/2})$ particle-($\pi g_{9/2})$ hole state to which  the ESPM can be applied.

With the  ingredients, as complete as feasible given above, one can  now proceed to  do the mapping for the $^{100}$Sn GT decay. Phrased in terms of the GT strength $\BGT$ defined in \cite{GSI,RIKEN}, the Landau Fermi-liquid fixed point  GT strength  can be equated to the GT strength given in terms of the ESPM quantities 
\be
{\BGT}^{\rm theory}={\cal{B}}_{\rm GT}^{\rm EPSM} (q_{\rm ssb} {q^{\rm Landau}_{\rm snc}})^2 \approx 10.8 q^2_{\rm ssb}\label{bgt}
\ee
with ${\cal{B}}_{\rm GT}^{\rm EPSM} =160/9$ and the full mixing factor given in the LFL fixed-point theory (\ref{Landau-snc}). This is the principal prediction of the theory that follows from the matching of the Fermi liquid to the doubly-magic shell structure.  Given  a well-measured experimental value ${\BGT}^{\rm exp}$, one could then extract the fundamental quenching factor $q_{\rm ssb}$ from (\ref{bgt}).

\section{Evidences} 
Let me first give an evidence for no $AIQ$, $q_{\rm ssb}\approx 1$. This would then -- with (\ref{Landau-snc}) -- give the prediction
\be
{\BGT}\approx 10.8.
\ee
The measurement made in GSI~\cite{GSI} that zeroes-in on $\sim 95\%$ of the daughter state of a pure $(\nu g_{7/2})$ particle-($\pi g_{9/2})$ hole configuration gives
\be
{\BGT}^{\rm GSI} \approx 10.\label{gsiresult}
 \ee
 This leads to no appreciable {\it FQ}.  This result is of course approximate, with $\sim 5\%$ deviation from the pure EPSM structure of the daughter state.  As discussed also in \cite{GSI}, there can be more refined analysis taking into account possible corrections to the EPSM but it cannot be entirely free of uncontrolled nuclear model dependence. That there is little, if any, evidence for $AIQ$ in this experiment is a robust observation. This result is globally consistent with the absence of any indication for $AIQ$ in other nuclear axial processes. I will come back to this matter below.
 
\subsection*{Evidence for big $AIQ$}

Now let's turn to the more recent, what's claimed to be ``improved," RIKEN result~\cite{RIKEN}, 
\be
{\BGT}^{\rm RIKEN}= 4.4^{+0.9}_{-0.7}.\label{RIKENdata}
\ee
This  is in serious tension with the GSI result, (\ref{gsiresult}). 
Within the range of values involved, this implies
\be
q_{\rm ssb}^{\rm RIKEN}\approx 0.58 - 0.69.\label{SSB}
\ee
This result implies that the ``fundamental" $g_A$ in the axial current effective  in nuclear medium defined by the chiral symmetry scale $\sim 4\pi f_\pi$ can be quenched from 1.276 to $\sim (0.74 - 0.88)$.  This applies to all axial transitions in nuclei. This is a big {\it fundamental} quenching.  




One immediate question to raise is  whether one cannot arrive at (\ref{RIKENdata}) by other mechanisms that do not require the $AIQ$. There may be several possibilities available in the literature but as far as the author is aware none seems viable as they stand. One notable case to illustrate the issue is the  {\it ab initio} calculation heralded as  a ``first-principles resolution" of the $g_A$ puzzle~\cite{firstprinciple}.  The strategy in this calculation -- which is anchored on standard nuclear chiral EFT -- is to ``tweak" the resolution scale (or effectively cut-off scale) in the EFT to shift the quenching effect from one-body to 2 or more-body currents to arrive at  $q_{\rm snc}$ that requires no $AIQ$ of the magnitude of (\ref{SSB}).
As argued in \cite{gA-MR}, this strategy exploiting  the N$^{3}$LO chiral expansion terms is untenable unless there is justification to ignore the next-order (N$^{4}$LO) terms.  As far as I can see, there is none given that the N$^{4}$LO terms will contain far too many unknown parameters.  This makes moot the assertion made in \cite{firstprinciple}.\footnote{What was obtained in \cite{firstprinciple}  corresponded to  (\ref{gsiresult}),  not to (\ref{RIKENdata}).} 

I now turn to another potentially serious indication for a possible $AIQ$ quenching in the recent developments on the spectral shape of multifold forbidden $\beta$ decay which turns out to be extremely sensitive to the $g_A$ coupling. Comparing theoretical predictions to the experimental $\beta$ spectrum, it has been argued~\cite{spectral} that 
the spectral shape of the multifold forbidden $\beta$ decay of $^{115}$In requires a quenching of $g_A$ by a factor $q^{^{115}{\rm In}} \sim (0.65 - 0.75)$ with the range of the value accounting for the nuclear model dependence in calculating the matrix elements of the weak current involved in the spectrum shape. Suppose the axial matrix elements in the spectral shape were calculated with sufficient accuracy. Then one could take the quenching factor found in the measurement to be the $AIQ$, $q^{^{115}{\rm In}}_{\rm ssb}\sim (0.65 - 0.75)$.  This is comparable to the $AIQ$  factor found in the RIKEN experiment.

There  is however a serious caveat in arriving at this result. As noted by the authors of \cite{spectral}, the spectral shape and decay do not match.
One may understand this conundrum given that the operators involved could be of drastically different structure. Unlike in the superallowed decay in $^{100}$Sn where $q_{\rm snc}$ and $q_{\rm ssb}$ are disentangled from each other thanks to the hidden scale symmetry governing in the doubly-magic-shell structure, the one-body axial-current operator figuring in the spectral shape -- with multi-body currents ignored --   does not enjoy any known symmetry protection. Its matrix element could therefore be highly nuclear-model dependent~\cite{spectral}.
 
Finally  let me make a few brief remarks on evidences that argue against a large $AIQ$ effect.

 That $g_{Ae}/g_A$ is near 1 in light nuclei but deviates from 1 in heavier nuclei  is not impossible,  but it is more or less ruled out in the  formula for $AIQ$ (\ref{aiq}) unless $\beta^\prime$ is big, say, $> 3$.  Although $q_{\rm ssb}$ cannot be calculated at present, there is nothing to suggest a big density dependence either. Furthermore $g_A^L\simeq 1$ seems to permeate from $\sim n_0$ to the dilaton-limit fixed point $\gsim 25 n_0$~\cite{MR-review}. 
 
 Another indication that an $AIQ$ factor of the magnitude (\ref{SSB}) is at odds with the quantity defined as $\epsilon_{\rm MEC}$ which zeroes in on the contribution from the strongly enhanced soft-pion exchange 2-body contribution to the matrix element of the axial charge operator $J_{\gamma_5 0}^i$ in first-forbidden $\beta$ transitions.  The enhancement factor $\epsilon_{\rm MEC}$  has the merit of being nuclear-model independent. First being protected by the soft-pion theorems, the exchange current is totally model-independent. Furthermore the ratio of the 2-body matrix element to the leading 1-body term is also nuclear model-independent. The measurements in Pb nuclei~\cite{warburton} showed indeed a strong enhancement in $\epsilon_{\rm MEC}$ predominantly controlled by chiral symmetry. The experimental  result of \cite{warburton} is precisely reproduced in the framework of G$n$EFT {\it without}  the $AIQ$ factor~\cite{KR}.  This  Pb result was also quantitatively supported in $A=12$ nuclei where the density dependence in $\Phi$ is consistent with the BR scaling~\cite{minamisono}. This -- what one might call a -- ``precision" test of chiral symmetry in nuclear systems would be ruined if $g_A$ were quenched by the fundamental renormalization (\ref{SSB}).  
 Of course in the context of the present day chiral effective field theory, focusing  on  the $\epsilon_{\rm MEC}$, both measured and calculated, could however raise the question as to the precision with which the one-body matrix element has  been calculated and whether all the terms in the weak current are taken into account. This matter could  be checked in modern high-powered {\it ab initio} calculations.

I should mention that the idea to exploit this enhanced  $J_{\gamma_5 0}^i$  in first-forbidden transitions - in a stark contrast to the highly suppressed 2-boy current for superallowed Gamow-Teller transitions -- was in fact studied  in combination with the effect on the spectral shape in \cite{suhonen-axial}. It would be interesting to pursue this direction in the context mentioned above.

\section{Concluding remarks}  In this note it was shown by mapping the RG-based Landau Fermi-liquid fixed point approximation to the ESPM in heavy doubly-magic nuclei that what's commonly referred to as ``quenched $g_A$," i.e., $g_A^O$, in the literature consists of two factors: one $q_{\rm snc}$ given entirely by ``snc" (strong nuclear correlations) and the other, $q_{\rm ssb}$ induced by the trace anomaly of QCD {\it which is invisible in the vacuum but exposed by density in nuclear medium.} It was also shown that $q_{\rm snc} g_A\to 1$ independent of density. While there is no established evidence in nuclear processes for the presence of a fundamental quenching $q_{\rm ssb}\neq 1$, there appear in the literature several experiments that  show an indication for  as much as $q_{\rm ssb}\sim 0.6$. {\it Such a big quenching in the fundamental coupling constant, if not ruled out,   would make an impact  not only in  nuclear theory in general but also in the effort to go beyond the SM, such as  $0\nu\beta\beta$ decay, in particular.} Such an important fundamental renormalization of $g_A$ in nuclear medium would pose a challenge to explain the apparent absence of visible effects in pion-nuclear dynamics.

I presented in this paper a suggestion for a well-defined resolution to the problem involving $g_A$: First establish $q_{\rm ssb}$ in the superallowed GT matrix element in the doubly-magic nucleus $^{100}$Sn, incorporate it in the analysis of the spectral shape of highly forbidden $\beta$ decays, and then obtain the wave functions that {\it fit} the spectral shape. One could then employ the resulting nuclear wave functions for the Gamow-Teller matrix elements in the $2\nu\beta\beta$ and $0\nu\beta\beta$ processes where non-negligible momentum transfer could be involved.

Given that $q_{\rm ssb}$ involves two unknown parameters, $\beta^\prime$ and $c_A$, it would require (at least) two superallowed GT transitions of doubly magic nuclei of different densities to extract the two unknowns. If there were another nucleus with density different from that of  $^{100}$Sn, that would offer the  information on $\beta^\prime$ in QCD that can be gotten only in baryonic matter.

\subsection*{Acknowledgments}

The author is grateful for useful and  informative comments from Jouni Suhonen.

 \end{document}